%Paper: hep-th/9407127
%From: YVONNE@urhep.pas.rochester.edu
%Date: Wed, 20 Jul 1994 09:39:58 -0500 (EST)

\magnification=1200
\baselineskip=13pt
\tolerance=100000
\overfullrule=0pt

\rightline{UR-1364\ \ \ \ \ \ }
\rightline{ER40685-814}

\bigskip

\baselineskip=18pt

\centerline{\bf NEW LINK INVARIANTS AND YANG-BAXTER EQUATION}

\medskip

\centerline{by}

\medskip

\centerline{Susumu Okubo}

\centerline{Department of Physics and Astronomy}

\centerline{University of Rochester}

\centerline{Rochester, NY 14627}

\vskip 2 truein

\noindent {\bf \underbar{Abstract}}

\medskip

We have new solutions to the Yang-Baxter equation, from which we have
constructed new link invariants containing
  more than two arbitrary parameters.  This
may be regarded as a generalization of the Jones'  polynomial.  We have
also found another simpler invariant which discriminates only the linking
structure of knots with each other, but not details of individual knot.

\vskip 2truein


\vfill\eject

\noindent {\bf 1. \underbar{Introduction}:}

\medskip

Knot or link invariants are useful to distinguish two topologically
inequivalent knots and links from each other.  The well-known examples are
those of Conway, Jones, Kauffman, and Homfly polynomials [1].  Although
these invariants can be constructed in a variety of ways, one interesting
method is to begin with solutions of Yang-Baxter equations ([1] and [2]).
The purpose of this note is to present some new knot and link invariants in
this manner.  We will show, first, in section 2, the existence of a link
invariant which distinguishes only linking structure but \underbar{not} the
individual knot configuration of each component knot contained in the link.
 The solution possesses as many arbitrary parameters as are desired so as to
enable us in general
  to distinguish any two linking structures.  In section 3, we will
consider a more general situation to obtain a  family of knot
invariants containing  two arbitrary integer parameters
by solving the Yang-Baxter equation (hereafter referred to as
YBE).  The new invariants may be considered as a generalization of the
one-parameter Jones' polynomials but differs from those of Kauffman and
Homfly's.

Since we start with the YBE in our construction, we will briefly sketch the
material relevant to our calculations.  Let $V$ be a finite-dimensional
vector space of dimension $N$, i.e.
$$N =\ {\rm Dim}\ V \quad . \eqno(1.1)$$
Let $e_1, e_2, \dots , e_N$ be a basis of $V$ and consider a linear mapping
$R (\theta)\ ;\  V \otimes V \rightarrow V \otimes V$ by
$$\underline{R} (\theta)\  e_a \otimes e_b = \sum^N_{c,d =1} R^{dc}_{ab}
(\theta) e_c \otimes e_d \eqno(1.2)$$
in terms of scattering matrix elements $R^{dc}_{ab} (\theta)$ where
  $\theta$ is
the spectral variable which may be identified as the rapidity, if we wish.
 Next,
set
$$V^n = V \otimes V \otimes \dots \otimes V \qquad (n{\rm -times})
\eqno(1.3)$$
and introduce $\underline{R}_{ij} (\theta) \ :\ V^n \rightarrow V^n \ (i,j
=
1,2, \dots, n, i < j)$ in the analogous fashion ([2] and [3]), which
operates only in the $i$-th and $j$-th vector spaces contained in the
tensor product $V^n$.  Then, the $\theta$-dependent YBE is the equation
$$\underline{R}_{12} (\theta) \underline{R}_{13} (\theta^\prime)
 \underline{R}_{23} (\theta^{\prime \prime}) =
 \underline{R}_{23} (\theta^{\prime \prime})  \underline{R}_{13}
(\theta^\prime )  \underline{R}_{12} (\theta)\eqno(1.4)$$
where variables $\theta,\ \theta^\prime,\ {\rm and}\ \theta^{\prime \prime}
$ satisfy the constraint
$$\theta^\prime = \theta + \theta^{\prime \prime} \quad . \eqno(1.5)$$
For our study of the knot and link invariants, the $\theta$-dependence is
actually superfluous, and we need consider only $\theta$-independent YBE:
$$ \underline{R}_{12}  \underline{R}_{13}  \underline{R}_{23} =
\underline{R}_{23}  \underline{R}_{13}  \underline{R}_{12} \quad .
\eqno(1.6)$$
Evidently, Eq. (1.6) may be regarded as a special case of Eq. (1.4) by
setting
$$ \underline{R}_{ij} =  \underline{R}_{ij} (\theta =0) \qquad {\rm or}
\qquad  \underline{R}_{ij} (\theta = \infty) \quad , \eqno(1.7)$$
provided that $ \underline{R}_{ij} (\theta)$ is not singular at $\theta =0$
and/or $\theta = \infty$.

Let  $\underline{P}_{ij} \ (i,j=1,2, \dots , n,i \not= j)$ : $V^n
\rightarrow V^n$ be the permutation operator of the $i$-th and $j$-th
vectors in $V^n$, and set
$$\sigma_j =  \underline{P}_{j,j+1}  \underline{R}_{j,j+1} \qquad (j=1,2,
\dots , n-1) \quad . \eqno(1.8)$$
Then, it is known ([2] and [3]) that the $\theta$-independent YBE (1.6)
will lead to
$$\sigma_{j+1} \sigma_j \sigma_{j+1} = \sigma_j \sigma_{j+1} \sigma_j
\qquad (j=1,2, \dots , n-2) \eqno(1.9)$$
in addition to
$$\sigma_j \sigma_k = \sigma_k \sigma_j \quad , \quad {\rm if}\quad
|j-k| \geq 2 \quad . \eqno(1.10)$$
Assuming hereafter that the inverse $\sigma^{-1}_j$ or equivalently
$ \underline{R}^{-1}_{ij}$ exists, then Eqs. (1.9) and (1.10) are precisely
the Artin's relations for the braid group $B_n$ of $n$-strings, which is
generated by 1, $\sigma_j$, and $\sigma_j^{-1}\ (j=1,2, \dots, n-1)$.

Any link can now be constructed out of braids in view of the Alexander
theorem ([1] and [4]) by identifying both ends of the strings in the braid.
 However, in order to construct a link invariant, we further assume the
existence of the Markov trace $\phi_n (g)$ in $B_n$ for
 $g \ \epsilon\ B_n$, which satisfies the Markov conditions:

\item{(i)} $\phi_n (gg^\prime) = \phi_n (g^\prime g) \quad , \quad (g,
g^\prime \ \epsilon \ B_n)$ \hfill (1.11a)

\item{(ii)} $\phi_{n+1} (g \sigma_n) = \tau \phi_n (g) \qquad (g\ \epsilon
\ B_n )$ \hfill (1.11b)

\item{(iii)} $\phi_{n+1} \left( g \sigma^{-1}_n \right) = \overline \tau
\phi_n (g) \qquad (g \ \epsilon\ B_n)\quad .$ \hfill (1.11c)

\noindent Here $\tau$ and $\overline \tau$ are some non-zero constants.
The link invariant associated with the Markov trace is now given by
$$M_n (g) = \left( {1 \over \tau \overline \tau}\right)^{{n-1 \over 2}}
\left( {\ \overline \tau\  \over \ \tau\ }
\right)^{{1 \over 2}\ w(g)} \phi_n (g)
\eqno(1.12)$$
where $w (g)$ is the exponent sum of the generators appearing in the
braid $g$ (for example, if $g = \sigma^3_1 \sigma^{-1}_2$, then
 $w(g) = 3 -1 = 2)$.  In this paper, we identify the Markov trace to be
$$\phi_n (g) = \ {\rm Tr}\ \rho (g) \quad . \eqno(1.13)$$
Here $\rho (g)$ is the representation matrix of $g \ \epsilon \ B_n$ in the
module $V^n$ on which $g$ acts.  The Markov conditions are then satisfied,
provided that the scattering matrix $R^{dc}_{ab}\  (\equiv R^{dc}_{ab} (
\theta = 0))$ obeys
$$\sum^N_{j=1} R^{dj}_{aj} = \tau \delta^d_a \quad , \quad
\sum^N_{j=1} (R^{-1})^{jc}_{jb} = \overline \tau \delta^c_b \quad .
\eqno(1.14)$$
We mention also the fact that we can always set $\tau = \overline \tau =1$
for all results given in sections 2 and 3 of this note.  We then have
$$M_n (g) = \phi_n (g) =\ {\rm Tr}\ \rho (g) \quad . \eqno(1.15)$$

\medskip

\noindent {\bf 2. \underbar{Multi-parameter Solution of YBE and
 Link Invariant}}

\medskip

Let $V$ be the $N$-dimensional vector space as in section 1, and consider
linear mappings $J_\mu \ :\ V \rightarrow V$ for $\mu = 1,2, \dots , p$,
satisfying
$$J_\mu J_\nu = J_\nu J_\mu \quad . \eqno(2.1)$$
Suppose that $\underline{R} (\theta)\ :\ V \otimes V \rightarrow V \otimes
V$ is given by
$$\underline{R} (\theta) x \otimes y = \sum^p_{\mu , \nu = 1} B_{\mu \nu}
(\theta) J
_\mu x \otimes J_\nu y \eqno(2.2)$$
for some functions $B_{\mu \nu} (\theta)$ of $\theta$.  Defining
 $\underline{R}_{ij} (\theta)$ similarly in $V^n$, it is trivial to see
$$\underline{R}_{ij} (\theta)  \underline{R}_{k \ell} (\theta^\prime ) =
 \underline{R}_{k \ell} (\theta^\prime )  \underline{R}_{ij} (\theta)
\quad . \eqno(2.3)$$
Therefore, Eq. (2.2) furnishes not only a solution of the YBE (1.4) but
also that of the classical YBE [3],
$$\left[  \underline{R}_{12} (\theta) ,  \underline{R}_{13} (\theta^\prime)
\right] + \left[  \underline{R}_{12} (\theta) ,  \underline{R}_{23}
(\theta^{\prime \prime} ) \right] + \left[  \underline{R}_{13}
(\theta^\prime ),  \underline{R}_{23} (\theta^{\prime \prime} \right]
= 0 \quad . \eqno(2.4)$$
Note that $B_{\mu \nu} (\theta)$ for $\mu, \nu = 1,2, \dots , p$ are
arbitrary functions of $\theta$.

In order to construct the link invariant, we set $\theta = \theta^\prime =
\theta^{\prime \prime} = 0$  with $ \underline{R} =  \underline{R}
(\theta =0)$ and $B_{\mu \nu} = B_{\mu \nu}
(\theta =0)$.  Expressing the operation of $ \underline{R}^{-1}$ in $V^2$
similarly by
$$ \underline{R}^{-1} x \otimes y = \sum^p_{\mu , \nu =1}
\overline B_{\mu \nu} J_\mu x \otimes J_\nu y \quad , \eqno(2.5)$$
the Markov conditions Eqs. (1.14) are rewritten now as
$$\eqalignno{&\sum^p_{\mu , \nu = 1} B_{\mu  \nu} J_\mu J_\nu = \tau \ {\rm
Id} \quad , &(2.6a)\cr
&\sum^p_{\mu , \nu = 1} \overline B_{\mu \nu} J_\mu J_\nu = \overline \tau
\ {\rm Id}&(2.6b)\cr}$$
where Id stands for the identity map in $V$.

A simple realization satisfying all these conditions is easily found, as
follows.  Suppose that we have

\item{(i)} $p=N$ \hfill (2.7a)

\item{(ii)} $J_\mu J_\nu = \delta_{\mu \nu} J_\mu \qquad\qquad
(\mu , \nu = 1,2, \dots , N)$ \hfill (2.7b)

\item{(iii)} $\sum^N_{\mu = 1} J_\mu = \ {\rm Id} \quad .$ \hfill (2.7c)

\noindent Note that $J_\mu$ may be identified with the projection operator
of the basis vector $e_\mu$ as
$$J_\mu e_\nu = \delta_{\mu \nu} e_\nu \eqno(2.8)$$
which we assume hereafter.  Because of $p=N$, both greek and latin indices
can now take the same range of values $1,2, \dots , N$, so that  we shall
hereafter in this section use them interchangeably.  Then, the scattering
matrix can be expressed as
$$R^{dc}_{ab} (\theta) = \delta^d_b \delta^c_a B_{ab} \quad , \quad
\left( R^{-1}\right)^{dc}_{ab} = \delta^d_b \delta^c_a \overline B_{ab}
\eqno(2.9)$$
with
$$\overline B_{ab} = \left( B_{ab} \right)^{-1} \quad . \eqno(2.10)$$
Assuming moreover,
$$B_{11} = B_{22} =
 \dots = B_{NN} = 1 \quad , \eqno(2.11)$$
the Markov conditions Eqs. (2.6) are satisfied with
$$\overline \tau = \tau = 1 \quad . \eqno(2.12)$$
Note that $B_{\mu \nu}$ for $\mu \not= \nu$ are completely arbitrary
constants as long as they are non-zero.

We can now compute the Markov invariant for any link, when we note
$${\rm Tr}\ J_\mu = 1 \quad . \eqno(2.13)$$
For  example, consider the link corresponding to the braid $g =
\sigma^2_1$ with $n=2$ which is depicted graphically in Fig. 1.  It is easy
to calculate
$${\rm Tr}\ \sigma^2_1 = \sum^N_{\mu , \nu=1}
B_{\mu \nu} B_{\nu \mu} \left( {\rm Tr}\ J_\mu \right) \left( {\rm Tr}\
J_\nu \right) = \sum^N_{\mu , \nu = 1} B_{\mu \nu} B_{\nu \mu} \quad ,
\eqno(2.14)$$
where we have written $\rho (\sigma_1) = \sigma_1$ for simplicity
 with the same convention hereafter.

\ \ \ \ \

\vskip 2.3 truein

\centerline{{\bf Figure 1.} A link corresponding to the braid $g =
\sigma^2_1$.}

\bigskip

\vfill\eject

\noindent In  Fig. 1, we designated two independent loops contained
therein as $\mu$ and $\nu$, respectively by reasoning to be explained.

We can compute other invariants in a similar
fashion.  However, there exists a simple graphical realization for
computations of the invariant as follows:
First, suppose that the link consists of $m$ interlocking
 loops $(m \leq n)$.  We name these loops as $\mu, \nu, \dots$ etc.
 with directions as in
Fig. 1, which can assume $N$ values $1,2,\dots , N$.  For each intersection
of the directed $\mu$-th and $\nu$-th loops, we assign a factor $B_{\mu \nu
}$ or $\overline B_{\nu \mu} = (B_{\nu \mu})^{-1}$, depending upon whether
the $\mu$-th loop at the left crosses the $\nu$-th loop at the right above
or below (see  Figures 2 and 3):

\ \ \ \ \ \

\vskip 2 truein

\centerline{{\bf Figure 2.} Crossing of the string $\mu$ above $\nu$.}

\bigskip

\ \ \ \

\vskip 2 truein

\centerline{{\bf Figure 3.} Crossing of the string $\mu$ below $\nu$.}

\bigskip

We then multiply all these factors and sum upon all loop indices $\mu,
\nu, \dots$ over the values $1,2, \dots, N$.  Finally, we assign a factor $
N$ for any unknot (i.e. an isolated simple circle) in the link if it
exists.  The rule immediately gives the result of Eq. (2.14) for Fig. 1.  As a
more complicated example, consider the link depicted in Fig. 4
corresponding to the braid $g = \sigma^2_1 \sigma^{-2}_2$ with $n=3$ to
find
$${\rm Tr}\ \left( \sigma^2_1 \sigma^{-2}_2 \right) = \sum^N_{\mu , \nu ,
\lambda = 1} B_{\mu \nu} B_{\nu \mu} \left( B_{\nu \lambda}\right)^{-1}
\left( B_{\lambda \nu}\right)^{-1} \quad . \eqno(2.15)$$

\ \ \ \ \

\vskip 4.2 truein

\centerline{{\bf Figure 4.} Link corresponding to the braid
 $g = \sigma^2_1 \sigma^{-2}_2$.}

\bigskip

\noindent Although
 we have found the rule on the basis of the Markov trace, we can directly
verify its invariances against the Reidemeister's 3 moves ([1] and [4]).
Especially, Eq. (2.11) guarantees the invariance under the 3rd
Reidemeister's move as we can observe from Fig. 5.

\vfill\eject

\ \ \ \

\vskip 1.8 truein

\centerline{{\bf Figure 5.} Invariance under the 3rd Reidemeister's move.}

\bigskip

\noindent We remark here that Fig. 5 is the graphical realization of the
Markov condition Eq. (1.14) for $\tau = \overline \tau = 1$.  Also, we need
not represent now the link in terms of the braid for the calculation,
although we will do so for the sake of illustration in this note.

We note that for a pure knot, we have only a single loop, and hence that we
have always the trivial result $\phi_n (g) = N$, no matter how complicated
the knot is.  This is because we have $B_{\mu \mu} = 1$.  For example,
consider a pure knot depicted in Fig. 6, corresponding to the braid $g=
\sigma^3_1$ where we calculate Tr $\sigma^3_1 = \sum^N_{\mu =1} (B_{\mu
\mu} )^3 = N$.

\ \ \ \

\vskip 2.2 truein

\centerline{{\bf Figure 6.} A pure knot corresponding to the braid
$g = \sigma^3_1$.}

\bigskip

In summary, the present
invariant is useful only for determining the global interlocking
 nature of the link, ignoring all details of individual knot
 structures contained therein.

Although we have considered a particular solution given by Eqs.
 (2.7) and (2.8), we can proceed similarly for more general
 solutions of Eqs. (2.1) and (2.6).  Nevertheless, the resulting
 Markov invariant can tell us again only about the interlocking
 link structure but not on the individual knot structure.
However, we will not go into the details.  In this connection, it
 may be worthwhile to make the following comment.  Suppose that
 again we assume $p=N$ but
$$J_\mu e_\nu = e_{\mu + \nu} \qquad (\mu , \nu = 1,2, \dots , N)
\eqno(2.16)$$
instead of Eq. (2.8), where we impose the cyclic condition
$$e_{\mu + N} = e_\mu \qquad (\mu = 1,2,\dots ,N) \eqno(2.17)$$
for the basis vectors.  We can now readily verify the validity of
Eq. (2.1) with $J_\mu J_\nu = J_{\mu + \nu}$ and
$J_N =$ Id.  The scattering matrix is now given by
$$R^{dc}_{ab} (\theta) = F^{d-b}_{a-c} (\theta) \eqno(2.18)$$
where we have set
$$F^{d-b}_{a-c} (\theta) = B_{c-a,d-b} (\theta)\eqno(2.19)$$
by extending the definition of $B_{\mu \nu} (\theta)$ to satisfy
$B_{\mu \pm N,\nu}(\theta) = B_{\mu , \nu \pm N}(\theta) =
B_{\mu , \nu}(\theta)$.  The fact that Eq. (2.18) gives a solution of the
YBE for arbitrary function $F^\mu_\nu (\theta) \ (-N < \mu , \nu < N)$ can
be directly verified also from the component-wise YBE:
$$\sum^N_{j,k,\ell =1} R^{jk}_{a_1 b_1} (\theta) R^{\ell a_2}_{kc_1}
(\theta^\prime) R^{c_2b_2}_{j
\ell}(\theta^{\prime \prime}) =
\sum^N_{j, k, \ell =1} R^{\ell  j}_{b_1 c_1} (\theta^{\prime \prime}) R^{c_
2 k}_{a_1 \ell} (\theta^\prime) R^{b_2 a_2}_{kj} (\theta)\eqno(2.20)$$
after some calculations.  We can construct Markov invariants on the basis
of the solution Eq. (2.18).  However, since a more general case will be
discussed in the next section, we will not go into detail.

\medskip

\noindent {\bf 3. \underbar{YBE as Triple Product and New Knot Invariants}}

\medskip

In order to find non-trivial knot invariants, we must discover more
general solutions of the YBE.  For this, it is more convenient to recast
the YBE as the triple product equation [5].  We will consider only the case
of the $\theta$-indpendent YBE for simplicity in the following.  Let
$<\cdot | \cdot >$ be a symmetric bilinear non-degenerate form in the
vector space $V$ and set
$$g_{jk} = g_{kj} =\  <e_j |e_k> \quad , \quad (j,k=1,2, \dots ,N)\quad .
\eqno(3.1)$$
We raise the indices in terms of the inverse tensor $g^{jk}$ as
$$e^j = \sum^N_{k=1} g^{jk} e_k \quad . \eqno(3.2)$$
Any $x\ \epsilon\ V$ can  then be expanded as
$$x = \sum^N_{j=1} <x|e_j>e^j = \sum^N_{j=1} e_j <e^j |x> \quad .
\eqno(3.3)$$
We now introduce two triple linear products $[x,y,z]$ and $[x,y,z]^*$ in  $V$
by
$$\eqalignno{\big[ e^c, e_a , e_b \big] &= \sum^N_{d=1} e_d R^{dc}_{ab}
\quad , &(3.4a)\cr
\big[ e^d, e_b , e_a \big]^* &= \sum^N_{c=1} R^{dc}_{ab} e_c
 &(3.4b)\cr}$$
so that we have
$$R^{dc}_{ab} =\  <e^d | [e^c, e_a, e_b ]>\  =\  < e^c |[e^d , e_b, e_a ]^*>
\quad . \eqno(3.5)$$
Now, the $\theta$-independent YBE  can then be shown to be rewritten
as the triple-product equation
$$\sum^N_{j=1} \left[ v, \left[ u,e_j,z \right], \left[ e^j ,x,y \right]
\right]^* = \sum^N_{j=1} \left[ u, \left[ v, e_j , x \right]^*,
\left[ e^j ,z,y \right]^* \right] \quad . \eqno(3.6)$$
As we may easily see, the choice $x= e_{a_1},\ y = e_{b_1},\
z = e_{c_1},\ u=e^{a_2},$ and $v=e^{c_2}$ in Eq. (3.6) will reproduce Eq.
(2.20) for $\theta = \theta^\prime =
\theta^{\prime \prime} = 0$.
  Also Eq. (3.5) will lead to a constraint equation
$$<u|[v,x,y]>\  =\  <v|[u,y,x]^*> \eqno(3.7)$$
in the basis-independent notation.  The relationship between the triple
products and the linear mapping $\underline{R}$ given in Eq. (1.2) is
easily found to be
$$\underline{R}\  x \otimes y = \sum^N_{j=1} e_j \otimes [e^j , x, y] =
\sum^N_{j=1}\  [e^j , y, x]^* \otimes e_j \quad . \eqno(3.8)$$
Especially, if we define
$$\underline{R}^* = \underline{P}_{12} \ \underline{R}\ \underline{P}_{12}
\quad , \eqno(3.9)$$
then we obtain the symmetrical relation of
$$\underline{R}^* x \otimes y = \sum^N_{j=1} e_j \otimes [e^j, x, y]^* =
\sum^N_{j=1}\  [e^j ,y,x] \otimes e_j \quad . \eqno(3.10)$$
Note that the condition $\underline{R}^* = \underline{R}$ is equivalent to
have $[z,x,y]^* = [z,x,y]$.

After these preparations, we seek solutions of Eq. (3.6) with the ansatz of

$$\eqalignno{[x,y,z] = \sum^p_{\mu , \nu=1} \big\{ &A_{\mu \nu} <y|
J_\nu z> J_\mu x + B_{\mu \nu} <x|J_\nu y> J_\mu z\cr
& +C_{\mu \nu} <z|J_\nu x>
J_\mu y\big\}&(3.11a)\cr
[x,y,z]^* = \sum^p_{\mu , \nu=1} \big\{ &A_{\mu \nu} <y| J_\nu z> J_\mu x+
B_{\nu \mu} <x|J_\nu y> J_\mu z\cr
& + C_{\nu \mu} <z,|J_\nu x>J_\mu y\big\}
&(3.11b)\cr}$$
for some linear mapping $J_\mu \ (\mu = 1,2,\dots,p)$ in $V$, where
$A_{\mu \nu},\ B_{\mu \nu},$ and $C_{\mu \nu}$ are some constants to be
determined.  The constraint Eq. (3.7) is satisfied by Eq. (3.11), provided
that we have
$$<x |J_\mu y>\  =\  < J_\mu x|y> \quad . \eqno(3.12)$$
The action of $\underline{R}$ in $V \otimes V$ can be obtained from Eq.
(3.8) to be
$$\eqalign{\underline{R}\  x \otimes y = \sum^p_{\mu , \nu=1} \bigg\{
&A_{\mu \nu}
<x| J_\nu y> \sum^N_{j=1} e_j
\otimes J_\mu e^j\cr
& + B_{\mu \nu} J_\nu x \otimes J_\mu y + C_{\mu \nu} J_\nu
y \otimes J_\mu x \bigg\} \quad .\cr} \eqno(3.13)$$
Comparing this with Eq. (2.2), we see that $B_{\mu \nu}$ here stands really
for $B_{\nu \mu}$ of section 2, or equivalently we are interchanging the
role of $\underline{R}$ and $\underline{R}^*$.  Defining the inverse
$\underline{R}^{-1}$ similarly by
$$\eqalign{\underline{R}^{-1} x \otimes y = \sum^p_{\mu , \nu=1} \bigg\{
&\overline{A}_{\mu \nu}
<x| J_\nu y> \sum^N_{j=1} e_j
\otimes J_\mu e^j\cr
& + \overline B_{\mu \nu} J_\nu x \otimes J_\mu y +
\overline C_{\mu \nu} J_\nu
y \otimes J_\mu x \bigg\} \quad ,\cr} \eqno(3.14)$$
the Markov condition is now equivalent to
$$\eqalignno{&\sum^p_{\mu , \nu =1} \big\{ (A_{\mu \nu} +B_{\mu \nu}) J_\mu
J_\nu + ({\rm Tr}\ J_\nu ) C_{\mu \nu} J_\mu \big\} = \tau\
{\rm Id} &(3.15a)\cr
&\sum^p_{\mu , \nu =1} \big\{ (\overline A_{\mu \nu} +
\overline B_{\nu \mu})
 J_\mu
J_\nu + ({\rm Tr}\ J_\nu )  \overline C_{\nu \mu} J_\mu \big\} =  \overline
\tau \
{\rm Id} &(3.15b)\cr}$$
where we have set
$${\rm Tr}\ J_\nu = \sum^N_{k=1} <e^k | J_\nu e_k> \quad . \eqno(3.16)$$

We have now to impose some algebraic relations among $J_\mu$'s.  We will
\underbar{not} consider, however, those given by Eqs. (2.7) amd (2.8) in
this note because of the following reason.  Suppose that we assume the
validity of Eqs. (2.7) and (2.8).  Then, we can find the following solution
of the YBE (1.6) or equivalently (3.6):
$$\eqalignno{A_{\mu \nu} &= 0 \quad , &(3.17a)\cr
\noalign{\vskip 4pt}%
C_{\mu \nu} &= C \theta (\mu -\nu ) = \cases{C \ ,\ &if $\quad \mu \geq \nu$\cr
0\ ,\ &if $\quad \mu < \nu$}\quad , &(3.17b)\cr
\noalign{\vskip 4pt}%
B_{\mu \nu} &= B\  {g_\mu \over g_\nu} - \bigg\{
{C \over 2} + B \pm \bigg[ \bigg( {C \over 2} \bigg)^2 + B^2
\bigg]^{{1 \over 2}} \bigg\} \ \delta_{\mu \nu} &(3.17c)\cr}$$
for arbitrary constants $B,\ C,$ and $g_\mu\ (\mu=1,2,
\dots,N)$.  Especially, if we choose $B=1,\ C=q - {1 \over q},$ and $g_\mu
=1$ for a parameter $q$, it will lead to the well-known solution
[6] of
$$R^{dc}_{ab} = \left( q- {1 \over q}\right) \delta^d_a
\delta^c_b \left[ \theta (a-b) - \delta_{ab} \right] +
\delta^c_a \delta^d_b \left[ q \delta_{ab} +
\left( 1- \delta_{ab} \right) \right] \quad . \eqno(3.18)$$
However, the Markov condition Eq. (1.14) is \underbar{not} satified by this
solution except for the trivial case of $q=1$.  In order to obtain link
invariant, we must resort then to a more elaborate graphical analysis based
upon the state model [6].  Unfortunately, the method
 does not appear to be readily
extended to the more general solution Eq. (3.17).

Instead of Eqs. (2.7) and (2.8), we will assume the following relations
among $J_\mu$'s: First, we extend the range of values for Greek indices
 $\mu,\ \nu$ etc. to all integers with periodicity conditions
$$J_{\mu \pm p} = J_\mu \quad , \quad A_{\mu \pm p , \nu} =
A_{\mu , \nu \pm p} = A_{\mu \nu} \eqno(3.19)$$
 and similarly for $B_{\mu \nu}$ and $C_{\mu \nu}$.  Next, we asume

\item{(i)} $J_\mu J_\nu = J_{\mu + \nu}$\hfill (3.20a)

\item{(ii)} $J_0 = J_p =$ Id \hfill (3.20b)

\item{(iii)} Tr $J_\mu = N\ \delta_{\mu , 0}$ \hfill (3.20c)

\noindent where we have set
$$\delta_{\mu \nu} = \cases{1\ ,\ &if $\quad \mu = \nu$ (mod $p$)\cr
0\ ,\ &otherwise\cr}\quad . \eqno(3.21)$$
We must then have
$$N = pm \eqno(3.22)$$
for another positive integer $m$ by the following reason.  Setting
$$P = {1 \over p}\ \sum^{p-1}_{\mu =0} J_\mu \quad ,$$
it is easy to see
$$P J_\mu = J_\mu P = P \quad , \quad P^2 = P \quad .$$
Especially, Tr $P = m$ must be a positive integer.  On the other side, we
calculate
$${\rm Tr}\ P = {1 \over p} \sum^{p-1}_{\mu =0} \ {\rm Tr}\
J_\mu = {1 \over p}\ N$$
which leads to the validity of Eq. (3.22).

The basis vectors $e_j\ (j=1,2,\dots,N)$ may be labelled now as
$$e_j = e_{\mu , A} \ (\mu = 1,2,\dots ,p, \ A = 1,2,\dots,m)
\eqno(3.23)$$
with
$$<e_{\mu , A}| e_{\nu , B}>\  = \delta_{\mu + \nu, 0}
\delta_{AB} \quad , \eqno(3.24)$$
on which $J_\mu$ acts as
$$J_\mu e_{\nu , A} = e_{\mu + \nu ,A} \quad . \eqno(3.25)$$
Note that these relations are then consistent with Eq. (3.12).  Also, if $m
=1$, then Eq. (3.25) will reproduce Eq. (2.16).

Now, we insert the expressions in Eqs. (3.11) to both sides of the YBE
(3.6) and use Eq. (3.3).  After some calculations, we then find
$$\eqalign{O &= \sum^N_{j=1} \big\{ \big[ v,[u,e_j ,z],[e^j ,x,y]\big]^*
- \big[ u,[v,e_j,x]^*,[e^j,z,y]^*\big]\big\}\cr
&= \sum^p_{\mu , \nu , \lambda =1} \big\{ K_1 <x|J_\lambda y><u|J_\mu
z>J_\nu v -
\hat K_1 <z|J_\lambda y> <v| J_\mu x> J_\nu u\cr
&\qquad\qquad\quad + K_2 <u|J_\lambda y><z|J_\mu x>J_\nu v - \hat K_2
 <v|J_\lambda y> <x|J_\mu z>J_\nu u\cr
&\qquad\qquad\quad + K_3 <u|J_\lambda x><z|J_\mu y>J_\nu v - \hat K_3
 <v|J_\lambda z> <x|J_\mu y>J_\nu u\cr
&\qquad\qquad\quad + K_4 <u|J_\lambda u><y|J_\mu z>J_\nu x - \hat K_4
<u|J_\lambda v> <y|J_\mu x>J_\nu z\cr
&\qquad\qquad\quad + K_5 <v|J_\lambda z><u|J_\mu y>J_\nu x - \hat K_5
 <u|J_\lambda x> <y|J_\mu v>J_\nu z\cr
&\qquad\qquad\quad + K_6 <v|J_\lambda y><u|J_\mu z>J_\nu x - \hat K_6
 <u|J_\lambda y> <v|J_\mu x>J_\nu z\cr
&\qquad\qquad\quad + K_7 <v|J_\lambda u><z|J_\mu x>J_\nu y\cr
&\qquad\qquad\quad + K_8 <v|J_\lambda x><u|J_\mu z>J_\nu y \big\}
\quad .\cr}\eqno(3.26)$$
Here, we have, for simplicity, suppressed the indices $\mu,$ $\nu$, and
$\lambda$ with
$$K_j \equiv K_{j,\mu \nu \lambda} \quad (j = 1,2,\dots,8) \quad .$$
We also note that the term proportional to
$<v|J_\lambda z><u|J_\lambda x> J_\nu y$ is absent, since it will result
only from $B_{\mu \nu}$ terms in accordance with the result of section 2.
The explicit values for $K_j$ are given by
$$\eqalignno{K_1 = &\sum^p_{\alpha , \beta , \gamma , \tau =1} \big\{
\delta_{\alpha + \beta + \gamma + \tau , \mu} A_{\nu \beta}
\big[ A_{\alpha \tau} A_{\gamma \lambda} + B_{\alpha \tau} A_{\gamma \lambda}
\big]\cr
&\qquad\qquad\quad + \delta_{\alpha + \beta + \gamma + \tau , \lambda}
A_{\nu \beta} \big[ C_{\alpha \mu} B_{\gamma \tau} +
C_{\alpha \mu} C_{\gamma \tau}\big]\cr
&\qquad\qquad\quad + \delta_{\alpha + \beta + \gamma + \tau , \nu}
 \big[ B_{\beta \tau} C_{\alpha \mu} +
C_{\beta \tau} C_{\alpha \mu}\big] A_{\gamma \lambda}\big\}\cr
+ &\sum^p_{\alpha , \beta , \gamma  =1} \big\{ N\
\delta_{\alpha + \beta + \gamma , 0} A_{\nu \beta}
 C_{\alpha \mu} A_{\gamma \lambda} - C_{\gamma, \mu-\beta} A_{\alpha ,
\lambda -\gamma} C_{\nu - \alpha , \beta}\big\}
&(3.27a)\cr
\noalign{\vskip 5pt}%
K_2 = &\sum^p_{\alpha , \beta , \gamma =1} \big\{ A_{\nu , \lambda -\gamma}
A_{\alpha , \mu -\beta} B_{\gamma -\alpha , \beta} + A_{\nu , \mu - \gamma}
B_{\alpha , \lambda - \beta} C_{\gamma -\alpha , \beta}\cr
&\qquad\qquad\quad - B_{\gamma , \lambda -\beta} A_{\alpha , \mu -\gamma}
C_{\nu -\alpha , \beta} \big\}&(3.27b)\cr
\noalign{\vskip 5pt}%
K_3 = &\sum^p_{\alpha , \beta , \gamma =1} \big\{ A_{\nu , \lambda -\gamma}
A_{\alpha , \mu -\beta} C_{\gamma -\alpha , \beta} + A_{\nu , \mu - \gamma}
B_{\alpha , \lambda - \beta} B_{\gamma -\alpha , \beta}\cr
&\qquad\qquad\quad - A_{\gamma - \alpha , \mu} B_{\alpha , \lambda -\beta}
B_{\nu -\gamma , \beta} - A_{\gamma - \beta , \mu} A_{\alpha ,
\lambda -\gamma} C_{\nu - \alpha , \beta} \big\}&(3.27c)\cr
\noalign{\vskip 5pt}%
K_4 = &\sum^p_{\alpha , \beta , \gamma =1} \big\{ B_{\beta , \nu -\gamma}
A_{\lambda - \beta,  \alpha} C_{\gamma , \mu - \alpha} - A_{\gamma - \alpha ,
 \mu}
A_{\lambda - \beta, \alpha} B_{\nu -\gamma , \beta}\cr
&\qquad\qquad\quad - A_{\gamma  -\beta , \mu} B_{\lambda -\gamma , \alpha}
C_{\nu -\alpha , \beta} \big\}&(3.27d)\cr
\noalign{\vskip 5pt}%
K_5 = &\sum^p_{\alpha , \beta , \gamma =1} \big\{ B_{\beta , \nu -\gamma}
B_{\lambda  -\beta , \alpha} C_{\gamma ,  \mu -  \alpha} -
B_{\gamma , \mu - \beta}
B_{\lambda - \gamma , \alpha} C_{\nu -\alpha , \beta}\big\}&(3.27e)\cr
\noalign{\vskip 5pt}%
K_6 = &\sum^p_{\alpha , \beta , \gamma =1} \big\{ B_{\beta , \nu -\gamma}
C_{\alpha  -\beta , \mu} C_{\gamma ,  \lambda -  \alpha} + C_{\beta
 , \nu - \alpha}
C_{\alpha - \gamma ,  \mu} B_{\lambda - \beta , \gamma}\cr
&\qquad\qquad\quad - C_{\gamma , \mu -\beta} B_{\lambda  -\gamma ,
\alpha}
C_{\nu -\alpha , \beta} \big\}&(3.27f)\cr
\noalign{\vskip 5pt}%
K_7 = &\sum^p_{\alpha , \beta , \gamma =1} \big\{ B_{\beta , \nu -\gamma}
A_{\lambda - \beta , \alpha} B_{\gamma , \mu - \alpha} - B_{\mu  - \alpha ,
\gamma}
A_{\lambda - \beta , \alpha} B_{\nu - \gamma ,  \beta} \big\}&(3.27g)\cr
\noalign{\vskip 5pt}%
K_8 = &\sum^p_{\alpha , \beta , \gamma =1} \big\{ B_{\beta , \nu -\gamma}
C_{\alpha  -\beta , \mu} B_{\gamma ,  \lambda -  \alpha} -
B_{\mu -\alpha , \gamma}
C_{\lambda , \alpha - \beta} B_{\nu -\gamma , \beta}\cr
&\qquad\qquad\quad + C_{\beta , \nu -\alpha} C_{\alpha  -\gamma , \mu}
C_{\lambda  -\beta , \gamma}
- C_{\gamma , \mu -\beta} C_{\lambda , \alpha -\gamma}
C_{\nu -\alpha , \beta} \big\} \quad .&(3.27h)\cr}$$
Moreover $\hat K_j\ (j = 1,2,\dots ,6)$ are the same expression as $K_j$
except for the interchange of
$$B_{\mu \nu} \leftrightarrow B_{\nu \mu} \quad , \quad C_{\mu \nu}
\leftrightarrow C_{\nu \mu} \quad . \eqno(3.28)$$
If we are interested in the $\theta$-dependent YBE (1.4), then the
expressions (3.26) and (3.27) are still valid, if we interpret the product
ABC in that order, for example, by $A(\theta^{\prime \prime})
B(\theta^\prime ) C(\theta)$ for $K_j$ and $A(\theta ) B(\theta^\prime )
C(\theta^{\prime \prime})$ for $\hat K_j$, respectively.  Actually we will
have $K_5 = \hat K_5 = 0$ for the present $\theta$-independent case because
of the following reason.  We change first $\alpha \rightarrow \nu - \alpha$,
$\beta \rightarrow \mu - \beta,$ and $\gamma \rightarrow \lambda - \gamma$
in the second term of $K_5$ and then let $\alpha \rightarrow \gamma
\rightarrow \beta \rightarrow \alpha$ to see the desired cancellation of
the first term.

The YBE (3.6) is now satisfied, provided that we have
$$\eqalignno{K_j &= \hat K_j = 0 \qquad (j = 1,2,3,4,6)&(3.29a)\cr
K_7 &= K_8 = 0 \quad . &(3.29b)\cr}$$
Although it is difficult to find the general solution of Eqs.
(3.29), we found some special solutions which further satisfy the Markov
condition (3.15) given now by
$$\eqalignno{&\sum^p_{\lambda =1} \big( A_{\lambda , \mu -\lambda} +
B_{\lambda , \mu -\lambda}\big) +
N \ C_{\mu , 0} = \tau \delta_{\mu ,0}&(3.30a)\cr
&\sum^p_{\lambda =1} \big( \overline A_{\lambda , \mu -\lambda} +
\overline B_{\lambda , \mu -\lambda}\big) +
N \ \overline C_{0, \mu} = \overline \tau \delta_{\mu ,0} \quad .
&(3.30b)\cr}$$
When we note
$$\sum^N_{j=1} J_\mu e_j \otimes J_\nu e^j = \sum^N_{j=1} e_j \otimes
 J_{\mu + \nu} e^j \quad ,$$
then the relation $\underline{R}\ \underline{R}^{-1} =$ Id can be expressed
as
$$\eqalignno{&N\ \sum^p_{\lambda =1} \overline A_{\lambda \nu} A_{\mu , -
\lambda} + \sum^p_{\alpha , \beta =1} \big\{ \overline A_{\beta \nu} \big[
B_{\alpha , \mu - (\alpha + \beta)} + C_{\alpha , \mu -(\alpha + \beta)}
\big]\cr
&\qquad \quad +\overline B_{\alpha , \nu -(\alpha +\beta)} A_{\mu \beta}
+ \overline C_{\alpha , \nu -(\alpha +\beta)} A_{\mu \beta}
\big\} = 0  &(3.31a)\cr
&\sum^p_{\alpha , \beta =1} \big\{ \overline B_{\mu -\alpha , \nu -\beta}
B_{\alpha \beta} + \overline C_{\nu -\beta , \mu -\alpha}
C_{\alpha \beta}\big\} = \delta_{\mu , 0} \delta_{\nu ,0}&(3.31b)\cr
&\sum^p_{\alpha , \beta =1} \big\{ \overline B_{\nu -\beta , \mu -\alpha}
C_{\alpha \beta} + \overline C_{\mu -\alpha , \nu -\beta}
B_{\alpha , \beta}\big\} = 0 \quad . &(3.31c)\cr}$$

We seek solutions of the YBE with the ansatz of
$$\eqalignno{A_{\mu \nu} &= \bigg( \delta_{\mu + \nu ,0} -
{1 \over p} \bigg) A + D \quad , &(3.32a)\cr
\overline A_{\mu \nu} &= \bigg( \delta_{\mu + \nu , 0} - {1 \over
p} \bigg) \overline A + \overline D \quad , &(3.32b)\cr
C_{\mu \nu} &= \bigg( \delta_{\mu + \nu , 0} - {1 \over p}
\bigg) C+F \quad ,
&(3.32c)\cr
\overline C_{\mu \nu} &= \bigg( \delta_{\mu + \nu ,0} - {1 \over p}\bigg)
\overline C + \overline F \quad , &(3.32d)\cr}$$
for some constants $A,\ \overline A,\ C ,\ \overline C,
\ D,\ \overline D,\ F,\ {\rm and}\ \overline F$.  Moreover, we impose the
condition
$$\eqalignno{&\sum^p_{\lambda =1}   B_{\lambda , \mu - \lambda} = G
\quad , &(3.33a)\cr
&\sum^p_{\lambda =1} \overline B_{\lambda , \mu -\lambda} =
\overline G\quad ,&(3.33b)\cr}$$
as well as
$$\sum^p_{\alpha , \beta = 1} \overline B_{\mu -\alpha , \nu -\alpha}
B_{\alpha \beta} = \delta_{\mu , 0} \delta_{\nu , 0} - {1 \over p}\
\delta_{\mu +\nu , 0} + G \overline G \eqno(3.33c)$$
for all $\mu,\nu = 1,2, \dots , p$.
A simple solution satisfying Eqs. (3.33) is for example given by
$$\eqalign{B_{\mu \nu} &= k \bigg[ \delta_{\mu , 0} \delta_{\nu , 0} -
{1 \over p}\ \delta_{\mu + \nu , 0} \bigg] + {1 \over p}\ G
\quad ,\cr
\overline B_{\mu \nu} &= {1 \over k}\ \bigg[ \delta_{\mu , 0} \delta_{\nu ,
0} - {1 \over p}\ \delta_{\mu +\nu ,0}\bigg] + {1 \over p}\
\overline G\cr}$$
for any constant $k$.

We have then found the following three solutions
 of the YBE.
First, all these solutions must satisfy the conditions:

$$\eqalignno{&p(A^2 +C^2) + NAC = 0 \quad , &(3.34a)\cr
&\tau = pA + NC = p (pD + NF +G) =-pC^2/A \quad , &(3.34b)\cr
&p^2 \overline AA = p^2 \overline CC = \overline \tau \tau
=1 \quad , &(3.34c)\cr
&GD = GF = 0 \quad . &(3.34d)\cr}$$
The rests of relations are given then by

\medskip

\noindent {\bf \underbar{Solution 1}}

\smallskip

$$\eqalignno{&G = 0\quad , \quad A = pD \quad , \quad C = pF \quad ,
&(3.35a)\cr
&\overline G = 0 \quad , \quad \overline A = p \overline D \quad , \quad
\overline C = p \overline F \quad , &(3.35b)\cr}$$

\medskip

\noindent {\bf \underbar{Solution 2}}

\smallskip

$$\eqalignno{&G = -C^2 /A \quad , \quad D = F = 0 \quad , &(3.36a)\cr
&\overline G = - (\overline C)^2 / \overline A \quad , \quad
\overline D = \overline F = 0 \quad , &(3.36b)\cr}$$

\vfill\eject

\medskip

\noindent {\bf \underbar{Solution 3}}

\smallskip

$$\eqalignno{&G = 0 \quad , \quad D = C^4/pA^3 \quad , \quad F = C^3
/p A^2 \quad , &(3.37a)\cr
&\overline G= 0 \quad , \quad \overline D = (\overline C)^4 /p
(\overline A)^3 \quad , \quad \overline F = (\overline C)^3
/p (\overline A)^2 \quad , &(3.37b)\cr}$$

The special case of $p=1$ (and hence $N=m$) is of some interest.  In that
case, we have $B_{\mu \nu} =0$ for solutions 1 and 3, while solution 2 will
reduce to a special case of the one given in section 2.  Consider the
solution 1 for $p=1$, where the scattering matrix is now given by
$$\eqalignno{R^{dc}_{ab} &= A\ g^{dc} g_{ab} + C\ \delta^d_a
\delta^c_b \quad , &(3.38a)\cr
\big( R^{-1}\big)^{dc}_{ab} &= {1 \over A}\ g^{dc} g_{ab} + {1
\over C}\ \delta^d_a \delta^c_b &(3.38b)\cr}$$
with
$${C \over A} + {A \over C} = -N \quad . \eqno(3.38c)$$
If we normalize $R^{dc}_{ab}$ by setting $AC =1$, then this reduces to the
solution given by Kauffman [7] who has also shown that the resulting knot
invariant corresponds to the Jones' polynomial.  This fact can be seen also
as follows.  It is more convenient to normalize $R^{dc}_{ab}$ now by
$\tau = \overline \tau = 1$ and hence $A = -C^2$.  In that case, we find
$${1 \over C^2}\ R^{dc}_{ab} - C^2 (R^{-1})^{dc}_{ab} =
\bigg( {1 \over C} - C \bigg) \delta^c_b \delta^d_a \eqno(3.39)$$
which is the generating relation for the Jones' polynomial.

The three solutions given by Eqs. (3.32)-(3.37) contain two arbitrary
integers $p$, and $m$ as well as many constants in $B_{\mu \nu}$, when we
use the normalization $\tau = \overline \tau =1$.  However, we will no
longer have the simple relation such as Eq. (3.39) for the general case.
The resulting link invariants are moreover rather complicated.  For
example, we calculate here Tr $(\sigma_1)^\ell$ for any positive integer
 $\ell$:
$$\eqalign{{\rm Tr}\ (\sigma_1)^\ell =\  &(p-1) (NA +pC)^\ell + (p-1) (N^2
- p^2) p^{\ell-2} C^\ell\cr
\noalign{\vskip 4pt}%
&+ p^\ell (ND +pF)^\ell + p^{2(\ell -1)}  (N^2 -p^2) F^\ell\cr
\noalign{\vskip 4pt}%
&+ {1 \over 2}\ [1-(-1)^\ell ] Np^{\ell -1} G^\ell + {1 \over
2}\ [1+(-1)^\ell ] N^2 Q_\ell \quad , \quad (\ell \geq 1)\cr}
\eqno(3.40)$$
while $Q_\ell$ for $\ell = 2s =$ even $\geq 2$ is given by
$$\eqalign{Q_\ell = \sum^p_{\mu_1 , \mu_2 \dots , \mu_s =1}
\  \sum^p_{\nu_1, \dots ,
\nu_s =1} &\delta_{\mu_1 + \mu_2 + \dots + \mu_s ,0}\
\delta_{\nu_1 + \nu_2 + \dots + \nu_s , 0}\cr
& \times
 K_{\mu_1 \nu_1} K_{\mu_2 \nu_2} \dots
K_{\mu_s \nu_s}\cr} \eqno(3.41a)$$
$$K_{\mu \nu} = \sum^p_{\alpha , \beta =1} B_{\mu - \alpha , \nu -
 \beta} B_{\beta \alpha} \quad . \eqno(3.41b)$$
Especially, we note that $B_{\mu \nu}$ terms in Eq. (3.40) will not
contribute for the case of $\ell =$ odd, corresponding to knots as in Fig.
6 ($\ell =3$).
The present link invariants differ from these of both Kauffman and Homfly
ploynomials.  Also its relationship to the 3-dimensional approach
 due to Witten
[8] is not obvious.

In ending this note, we remark that we can find more solutions of YBE.  One
example is given by
$$A_{\mu , \nu} = D \quad , \quad C_{\mu , \nu} = F \quad , \quad \overline
A_{\mu \nu} = \overline D \quad , \quad \overline C_{\mu \nu} =
\overline F$$
satisfying conditions
$$\eqalign{&p^4 \overline DD = p^4 \overline FF = \overline \tau \tau
 = 1 \quad ,\cr
&p(D^2 + F^2) + NFD = 0 \quad ,\cr
&\tau = p (pD + NF )= -p^2 F^2/D\cr}$$
while $B_{\mu \nu}$ must obey relations
$$\eqalign{&\sum^p_{\lambda =1} B_{\lambda , \mu - \lambda} = \tau \bigg(
\delta_{\mu , 0} - {1 \over p}\bigg)\cr
&\sum^p_{\lambda =1} \overline B_{\lambda , \mu - \lambda} =
\overline \tau \bigg(
\delta_{\mu , 0} - {1 \over p}\bigg)\cr
&\sum^p_{\alpha , \beta =1} \overline B_{\mu - \alpha , \nu - \beta}
 B_{\alpha \beta} =
\delta_{\mu , 0} \delta_{\nu , 0}  - {1 \over p^2} \quad .\cr}$$
A solution for $B_{\mu , \nu}$ and $\overline B_{\mu , \nu}$
 satisfying these conditions is easily
found to be
$$\eqalign{B_{\mu \nu} &= \tau \bigg( \delta_{\mu , 0} \delta_{\nu , 0} - {
1 \over p^2} \bigg) \quad ,\cr
\overline B_{\mu \nu} &= \overline \tau \bigg( \delta_{\mu , 0}
 \delta_{\nu , 0} - {
1 \over p^2} \bigg) \quad .\cr}$$
There are other solutions in which we have $A_{\mu \nu} =0$.
However, the Markov conditions are satisfied only
for the rather uninteresting case of $p=N$.

\medskip

\noindent {\bf \underbar{Acknowlegement}}

\medskip

This work is supported in part by the U.S. Department of Energy
 Grant No. DE-FG-~02-91ER40685.

\vfill\eject

\noindent {\bf \underbar{References}}

\medskip

\item{1.} L.H. Kauffman, \underbar{Knots and Physics} (World Scientific,
Singapore-New Jersey-London-Hong Kong, 1991).

\item{2.} C.N. Yang~and M.L. Ge, (editors), \underbar{Braid Groups, Knot
Theory and Statistical} \underbar{Mechanics
 I. and II.}, (World Scientific,
Singapore-New Jersey-London-Hong Kong, I 1989 and II 1994).

\item{3.} M. Jimbo (editor), \underbar{Yang-Baxter Equation in
 Integrable Models}
(World Scientific, Singapore-New Jersey-London-Hong Kong, 1989).

\item{4.} C. Livingston, \underbar{Knot Theory} (The Mathematical
Association of America, Washington, 1993).

\item{5.} S. Okubo, Triple products and Yang-Baxter equation: I, Octonionic
and quaternionic triple systems: II, Orthogonal and symplectic ternary
systems,
  Jour. Math. Phys. {\bf 34} (1993) I. 3273 and II. 3292.

\item{6.} L.H. Kauffman, ref. [1], see pp.161-173.

\item{7.} L.H. Kauffman, ref. [1], see pp.111-116.

\item{8.} E. Witten,
Quantum field theory and the Jones' polynomial, Comm. Math. Phys. {\bf 121}
 (1989) 351.

\end